\documentclass[letterpaper,10pt]{article}

\usepackage{amsfonts}
\usepackage{fixmath}
\usepackage{setspace}
\usepackage{enumitem}
\usepackage{amsmath}
\usepackage{xcolor}
\usepackage{mathtools,url}
\usepackage{latexsym}
\usepackage{subcaption} 
\usepackage[font={small,sf}]{caption}
\usepackage{algorithm}
\usepackage[noend]{algpseudocode}

\newcounter{myprot}
\newcounter{myalg}
\newcounter{mythm}
\newenvironment{mythm}
{\refstepcounter{mythm}  \noindent \textbf{THEOREM \arabic{mythm}:}\em}
{\vspace{.25em}}

\newcounter{mylem}
\newcounter{mycor}
\newcounter{myobs}
\newcounter{mydef}
\newcounter{myconj}
\newenvironment{myproof}
{\noindent \textbf{PROOF:}}
{\vspace{-3ex}\begin{flushright} $\Box$ \end{flushright}\vspace{2ex}}

\makeatletter
\g@addto@macro{\UrlBreaks}{\UrlOrds}
\makeatother

\renewcommand{\footnotesize}{\fontsize{8pt}{10pt}\selectfont}

\newif\ifinappendix
\let\oldappendix\appendix
\renewcommand{\appendix}{
  \oldappendix
  \inappendixtrue
}

\usepackage{booktabs}
\usepackage[hidelinks]{hyperref}
\usepackage{xfrac}
\usepackage{wrapfig,multirow,relsize}
\usepackage[nospace,compress]{cite}
\usepackage{microtype}

\newcommand{\OmitText}[1]{ {} }

\usepackage{xspace}

\usepackage[all]{hypcap}

\usepackage[capitalise,nameinlink]{cleveref}
\crefname{section}{Sect.}{Sect.}
\Crefname{section}{Section}{Sections}

\usepackage{url}
\makeatletter
\g@addto@macro{\UrlBreaks}{\UrlOrds}
\makeatother
\makeatletter
\def\Url@twoslashes{\mathchar`\/\@ifnextchar/{\kern-.2em}{}}
\g@addto@macro\UrlSpecials{\do\/{\Url@twoslashes}}
\makeatother

\begin{document}

\title{\Large An Algorithm for Bounding the Probability of $r$-core Formation in $k$-uniform Random Hypergraphs}

\author{George Bissias \\ gbiss@cs.umass.edu \\ College of Information and Computer Sciences, UMass Amherst}

\maketitle

\setlength{\belowdisplayskip}{3pt} 
\setlength{\belowdisplayshortskip}{3pt}
\setlength{\abovedisplayskip}{3pt} 
\setlength{\abovedisplayshortskip}{3pt}


\section{Introduction}

We present an algorithm for bounding the probability of $r$-core formation in $k$-uniform hypergraphs. Understanding the probability of core formation is useful in numerous applications including bounds on the failure rate of Invertible Bloom Lookup Tables (IBLTs)~\cite{goodrich:2011} and the probability that a boolean formula is satisfiable~\cite{Molloy:2004}. 

\subsection{Problem Statement}

Let $\mathcal{H}^k_{v, p}$ be a $k$-uniform hypergraph on $v$ vertices $V$ and edge set $E$ where each of the $\binom{v}{k}$ edges in $E$ occurs with probability $p$. An $r$-core over vertices $U \subseteq V$ is an induced subgraph $\mathcal{H}^k_{v, p}(U)$ in which every vertex has degree at least $r$. Define $\mathcal{C}(v,p,k,r)$ to be the probability that at least one $r$-core forms in $\mathcal{H}^k_{v, p}$. In this paper, we seek upper and lower bounds on $\mathcal{C}(v,p,k,r)$.

\subsection{Establishing $r$-core Existence}

The succinct description of an $r$-core belies the complexity associated with identifying them. The standard approach is by means of a \emph{peeling} process~\cite{Molloy:2004}. Proceeding in rounds, all vertices with degree less than $r$ are removed from $\mathcal{H}^k_{v, p}$ along with any incident edges. The process is repeated in the subsequent rounds, terminating after the round where no vertices are removed. The remaining vertices form an $r$-core in $\mathcal{H}^k_{v, p}$.

\section{General Algorithm}

Calculating $\mathcal{C}_*(v,p,k,r)$, the probability that exactly one $r$-core forms \emph{anywhere} in $\mathcal{H}^k_{v, p}$ is relatively straightforward given knowledge of $\mathcal{C}_{\bar{u}}(v,p,k,r)$, the probability that an $r$-core forms on all vertices in some subset of $u$ vertices. Algorithm~\ref{alg:general} (appearing in Appendix~\ref{sec:gen_alg}) describes how to calculate $\mathcal{C}_*(v,p,k,r)$ from $\mathcal{C}_{\bar{u}}(v,p,k,r)$, which we prove correct in the remainder of this section. Given that $\mathcal{C}_{\bar{u}}(v,p,k,r)$ is known for all $u \leq v$, our approach is to recursively calculate $\mathcal{C}_u(v,p,k,r)$ --- the probability that a single $r$-core of size $u$, and no other, forms somewhere in $\mathcal{H}^k_{v, p}$ --- and use the partial results to construct $\mathcal{C}_*(v,p,k,r)$. Notice that the probability that $i$ $r$-cores form in $\mathcal{H}^k_{v, p}$ is bounded above by $\mathcal{C}_*(v,p,k,r)^i$. This is true because the latter quantity is equivalent to allowing edges to be reused between $r$-cores. From this it follows that
\[
\begin{array}{rl}
\mathcal{C}(v,p,k,r) & \leq \sum\limits_{i=1}^{\infty} \mathcal{C}_*(v,p,k,r)^i \\
& = \frac{\mathcal{C}_*(v,p,k,r)}{1 - \mathcal{C}_*(v,p,k,r)},
\end{array}
\] 
where the second step is the closed-form for a geometric series. 

We next show how $\mathcal{C}_{\bar{u}}(v,p,k,r)$ is related to $\mathcal{C}_{u}(v,p,k,r)$, the most critical step in the algorithm. 
Because we are working with random hypergraphs, the probability that an $r$-core forms on vertex set $U$ is equivalent to the probability that an $r$-forms on \emph{any other} set $U' \subseteq V$ where $|U'| = |U|$. Therefore $\mathcal{C}_{\bar{u}}(v,p,k,r) = \mathcal{C}_U(v,p,k,r)$, where the latter quantity is the probability that an $r$-core forms on all vertices in $U \subseteq V$. 
Because $\mathcal{H}^k_{v, p}$ is $k$-uniform, it is clear that no $r$-core can form when $v < k$. Hence,  $\forall u < k, \mathcal{C}_u(v,p,k,r) = 0$. The case where $u \geq k$ is covered by the following theorem.

\bigskip

\begin{mythm}
\label{thm:main}
For every $u$, $k \leq u \leq v$
\[
\mathcal{C}_u(v,p,k,r) = \mathcal{C}'(v,p,k,r) \mathcal{C}''(v,p,k,r)
\] 
where
\[
\mathcal{C}'_u(v,p,k,r) = \sum\limits_{U \subseteq V, u=|U|}  \binom{v}{u}  \mathcal{C}_U(v,p,k,r) \prod\limits_{x > u}^v \left(1 - \mathcal{C}_x(v,p,k,r) \right)^{\binom{v - u}{x - v}},
\]
and
\[
\mathcal{C}''_u(v,p,k,r) = 1 - \sum\limits_{x=k}^{v-u}  \mathcal{C}_{v-u}(x,p,k,r).
\]

\end{mythm}

\begin{myproof}
We proceed by showing that $\mathcal{C}'_u(v,p,k,r)$ is the probability that exactly one $r$-core of size $u$ forms somewhere in $\mathcal{H}^k_{v, p}$ and is not contained in any larger $r$-core. While $\mathcal{C}''_u(v,p,k,r)$ is the probability that no other $r$-core forms in $\mathcal{H}^k_{v, p}$ over a subset of vertices distinct from the set of vertices containing the $r$-core of size $u$. If both of these hold, then it is clear that $\mathcal{C}_u(v,p,k,r) = \mathcal{C}'_u(v,p,k,r) \mathcal{C}''_u(v,p,k,r)$, thus the theorem will be proved.

For $\mathcal{C}'_u(v,p,k,r)$, we begin by summing (over all possible $U \subseteq V$) the probabilities $\mathcal{C}_U(v,p,k,r)$ qualified by the restriction that $U$ not be contained in a larger $r$-core over vertices $X$. Let $u = |U|$. For any $x > u$, there exist $s = \binom{v-u}{x-u}$ supersets of size $x$ containing the $u$ vertices. This implies that the probability that the $r$-core on $u$ vertices \emph{is not} contained in any $r$-core of size $x$, is given by $(1 - \mathcal{C}_x(v,p,k,r))^s$, which can be extended by conjunction to all $x > u$. That is to say, the probability that an $r$-core forms over all vertices $U$ and that this $r$-core is a subset of no other $r$-core is given by
\[
\mathcal{C}_U(v,p,k,r) \prod\limits_{x > u}^v \left(1 - \mathcal{C}_x(v,p,k,r) \right)^{\binom{v - u}{x - v}}.
\]
Summing over the $\binom{v}{u}$ ways to choose a subset of $u$ vertices we arrive at the desired result, $\mathcal{C}'_u(v,p,k,r)$.

Turning to $\mathcal{C}''_u(v,p,k,r)$, we know that any $r$-core of size greater than $v-u$ must intersect every $r$-core of size $u$, which implies that it could not be a distinct $r$-core. Thus, we consider only distinct $r$-cores that form over $x$ vertices in the range $[k, v-u]$. Assuming that one $r$-core of size $u$ exists, a distinct $r$-core could only form in an induced subgraph of size $v-u$, which occurs with probability  $\mathcal{C}_{v-u}(x,p,k,r)$. It follows that the probability that \emph{no} distinct $r$-core of size $x$ forms, for any possible value of $x$, is given by $\mathcal{C}''_u(v,p,k,r)$. 
\end{myproof}

\section{Bounding Local $r$-core Formation}

With Theorem~\ref{thm:main} in hand, we next seek to measure $\mathcal{C}_U(v,p,k,r)$, the probability that an $r$-core forms over a specific subset of vertices $U$ in $\mathcal{H}^k_{v, p}$. This paper describes two different approaches: one gives upper and lower bounds based on hypergraph connectivity and another provides a close approximation based on vertex \emph{covering}.

\subsection{Connectivity Bound}

We begin with the observation that a connected component on vertices $U$ is equivalent to a 1-core on $U$. Thus, $\mathcal{C}_U(v,p,k,1)$ is the probability that the induced subgraph on $U$ is connected. Expanding on this idea, consider an \emph{interleaved} graph construction / peeling process yielding a hypergraph $\mathcal{I}^k_{v, p}$ wherein an $r$-core is revealed by peeling in rounds, one 1-core at a time, and edges are regenerated at random (removing the remaining old ones and adding new ones) with probability $p$ after each round. Let $\mathcal{C}^*_U(v,p,k,r)$ denote the probability that any $r$-core forms over all vertices $U$ in $\mathcal{I}^k_{v, p}$ using this interleaving process. In this alternative construction, an $r$-core on vertex set $U$ exists iff a 1-core on $U$ exists during each round. Thus, $\mathcal{C}^*_U(v,p,k,r) = \mathcal{C}_U(v,p,k,1)^r$. Moreover, as the next theorem shows, the probability $\mathcal{C}^*(v,p,k,r)$ that any $r$-core forms in $\mathcal{I}^k_{v, p}$ can be used to bound $\mathcal{C}(v,p,k,r)$ both above and below. 

\bigskip

\begin{mythm}
\label{thm:upper_lower}
In expectation, $\mathcal{C}^*(v,p/r,k,r) \leq \mathcal{C}(v,p,k,r) \leq \mathcal{C}^*(v,p,k,r)$.
\end{mythm}

\bigskip

\begin{myproof}
To show $\mathcal{C}^*(v,p/r,k,r) \leq \mathcal{C}(v,p,k,r)$, let $N$ be the expected number of edges formed in $\mathcal{H}^k_{v, p}$. For $\mathcal{I}^k_{v, p}$, we divide the number of edges uniformly between rounds in parcels of $N/r$. Since edges are cleared between rounds and their total number is $N$, it follows that the probability that an $r$-core forms in $\mathcal{I}^k_{v, p}$ with edge probability $p/r$ cannot exceed the probability that an $r$-core forms in $\mathcal{H}^k_{v, p}$ where all edges contribute simultaneously to the formation of an $r$-core. The inequality $\mathcal{C}(v,p,k,r) \leq \mathcal{C}^*(v,p,k,r)$ can be argued similarly by noting that $\mathcal{I}^k_{v, p}$ generates and clears $N$ edges in expectation \emph{per round}, thus the probability that an $r$-core develops in it cannot be less than the probability that one forms in $\mathcal{H}^k_{v, p}$.
\end{myproof}

\subsubsection{A recursive formula for connectivity probability}

Gilbert~\cite{Gilbert:1959} introduced the following classical result that gives the exact probability that a specific $u$ vertices in $\mathcal{H}^2_{v,p}$, an Erdos-Renyi random graph, are connected.
\[
f^2_p (u) = 1 - \sum\limits_{i=1}^{u-1} f^2_p(i) \binom{u-1}{i-1} (1-p)^{i(u-i)},
\]
with $f^2_p(1) = 1$. Function $f^2_p(v)$ can equivalently be interpreted as the probability that the entire graph $\mathcal{H}^2_{u,p}$ is connected. 

We next prove a more general result for $k$-uniform hypergraphs, beginning with the following definitions. For all $u \geq 1$,
\[
f^k_p (u) = 1 - \sum\limits_{i=1}^{u-1} f^k_p(u, i),
\]
where $f^k(1) = 1$, $f^k(u) = 0$ when $1 < u < k$,
\[
f^k_p(u, i) = f^k_p(i) \binom{u-1}{i-1} (1-p)^{\varepsilon^k(u,i)},
\] 
for $u \geq k$ and $1 \leq i < u$, and
\[
\varepsilon^k(u,i) = \sum_{j=1}^{\min(i, k-1)} \binom{i}{j} \binom{u-i}{k-j}.
\]

\bigskip

\begin{mythm}
\label{thm:conn_prob}
The probability that a certain set of $u$ vertices form a connected component in $\mathcal{H}^k_{v,p}$, $u \leq v$, is given by $f^k_p (u)$. Furthermore, $f^k_p(u, i)$, where $u \geq k$ and $1 \leq i < u$, gives the probability that there exists at least one set of $i$ vertices in $\mathcal{H}^k_{u,p}$ that connect to each other and to no others.
\end{mythm}

\bigskip

\begin{myproof}

We proceed by induction on $u$ as follows. Clearly $f^k_p(1) = 1$ is correct since every vertex forms a connected component with itself. And for $1 < u < k$, it is also clear that $f^k_p(u)$ should have value 0, because no edge forms on fewer than $k$ vertices. Now suppose that $f^k_p(u-1)$ gives the correct probability that a connected component of size $u-1$ forms in $\mathcal{H}^k_{v,p}$ when $u-1 \geq k$. The probability that a set of $u$ vertices is connected is equivalent to the complement of the probability that there exists some connected component that forms exclusively on a proper subset of those vertices, which is equal to $1 - \sum_{i=1}^{u-1} f^k_p(u, i)$ provided that $f^k_p(u, i)$ gives the indicated probability. Thus, it remains only to prove the correctness of our expression for $f^k_p(u, i)$, given that $f^k_p(i)$ holds for $i < u$, and the validity of $f^k_p(u)$ will follow. 

Suppose that the induction on $f^k_p(i)$ holds for components up to size $i = u-1$, and consider adding a new vertex $x$ to the component. We claim that $f(u, i)$ gives the probability that there exists a component on exactly $i$ vertices where $1 \leq i < u$. There are $\binom{u-1}{i-1}$ ways to choose a candidate subset $S$ of $i-1$ vertices from the original set of $u-1$ vertices. Adding vertex $x$ to $S$, means the candidate subset has $i$ vertices. Since $i < u$, we know by induction that all vertices in $S$ are connected with probability $f^k_p(i)$. So taken together, the probability of forming a connected component with size \emph{at least} $i$ is given by $f^k_p(i) \binom{u-1}{i-1}$. Now in order for candidate component $S$ to have \emph{exactly} $i$ vertices, it must be the case that none of its $i$ vertices are connected to any of the remaining $u-i$ vertices. In a $k$-uniform hypergraph, there are $\varepsilon^k(u,i)$ possible edges between vertices $S$ and the remaining vertices. And the probability that none of those edges exist is equal to $(1-p)^{\varepsilon^k(u,i)}$, which completes the proof.

\end{myproof}

\subsection{Covering Heuristic}

Another approach to measuring $\mathcal{C}_U(v,p,k,r)$, is to determine the probability that every vertex in $U$ is covered by at least $r$ edges, which is almost identical to $\mathcal{C}_U(v,p,k,r)$, except that it admits the additional possibility that multiple $r$-cores create a disjoint covering of $U$. In this section, we develop a function $\tilde{\mathcal{C}}_U(v,p,k,r)$, that closely approximates the coverage probability, and $\mathcal{C}_U(v,p,k,r)$ accordingly. 

For $\mathcal{H}^k_{v, p}$, we can imagine that edges in $E$ are formed as follows. Each edge has $k$ \emph{slots}, every slot can accommodate exactly one vertex from $V$, and no vertex can occupy more than one slot in a single edge. Define $b(x; n, p)$ and $B(x; n, p)$ to be the probability mass and cumulative distribution functions of the distribution $\texttt{Binomial}[n, p]$, respectively. Suppose that there are $e_U$ edges in $U$ with $u = |U|$. Placing vertices into slots independently at random defines a Poisson process. In particular, the probability that a given vertex is assigned to at least $r$ slots, independent of the other vertices is given by $1 - B(r-1; e_U, k/u)$. The actual number of edges in $\mathcal{H}^k_{v, p}(U)$ varies according to a separate Poisson process. There are $\binom{u}{k}$ possible edges in $U$, and each is present with probability $p$. Thus, the probability that exactly $e_U$ edges form within $\mathcal{H}^k_{v, p}(U)$ is equal to $b(e_U; \binom{u}{k}, p)$. With probabilities for edge and $r$-core formation in hand, we can now derive our approximation to $\mathcal{C}_U(v,p,k,r)$.
\begin{equation}
\label{eqn:covering}
\tilde{\mathcal{C}}_U(v,p,k,r) = \sum\limits_{e_U=0}^{\binom{u}{k}} b\left( e_U; \binom{u}{k}, p \right) (1 - B(r-1; e_U, k/u)).
\end{equation}

\section{Evaluation}

In this section we empirically investigate the accuracy of the local and global bounds provided in previous sections. Although the bounds we have presented apply to global $r$-core formation for any $r \geq 2$, our experiments focus exclusively on the case where $r = 2$. Overall, the connectivity bound on local 1-core formation is demonstrated to closely match actual probabilities generated via Monte Carlo (MC) trials. And as a result, compared to MC, it provides good upper and lower bounds on global 2-core formation by way of Algorithm~\ref{alg:general} and Theorem~\ref{thm:upper_lower}. Unfortunately, this bound also becomes numerically unstable as $v$, the number of vertices in $\mathcal{H}^k_{v, p}$, grows. For larger values of $v$, we show empirically that the covering heuristic can be used to gain a very good approximation to the probability of 2-core formation, even though it does not always provide a strict upper bound.

\begin{figure}
\centering
\begin{subfigure}{.5\textwidth}
  \centering
  \includegraphics[width=1.0\linewidth]{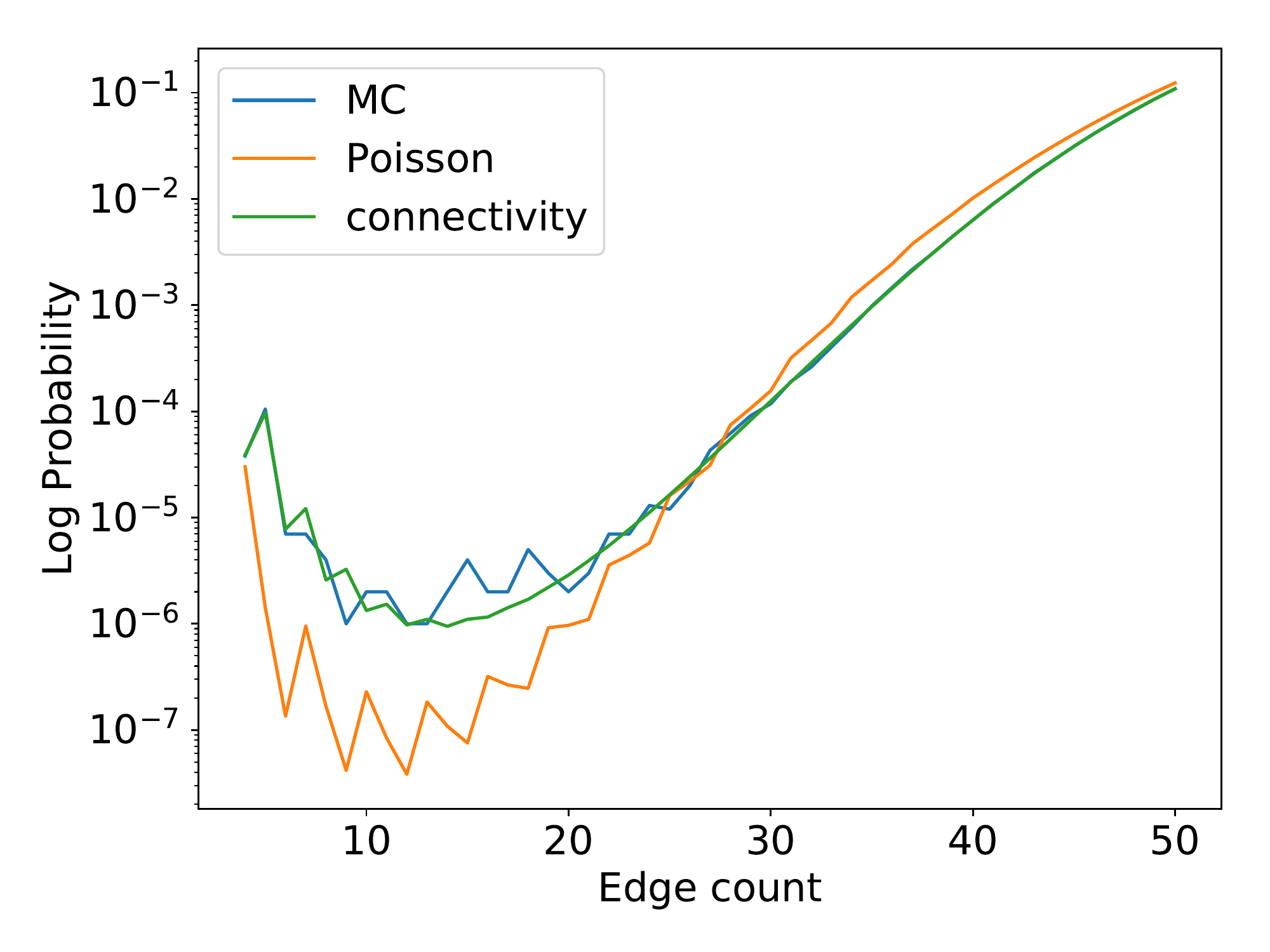}
\end{subfigure}%
\begin{subfigure}{.5\textwidth}
  \centering
  \includegraphics[width=1.0\linewidth]{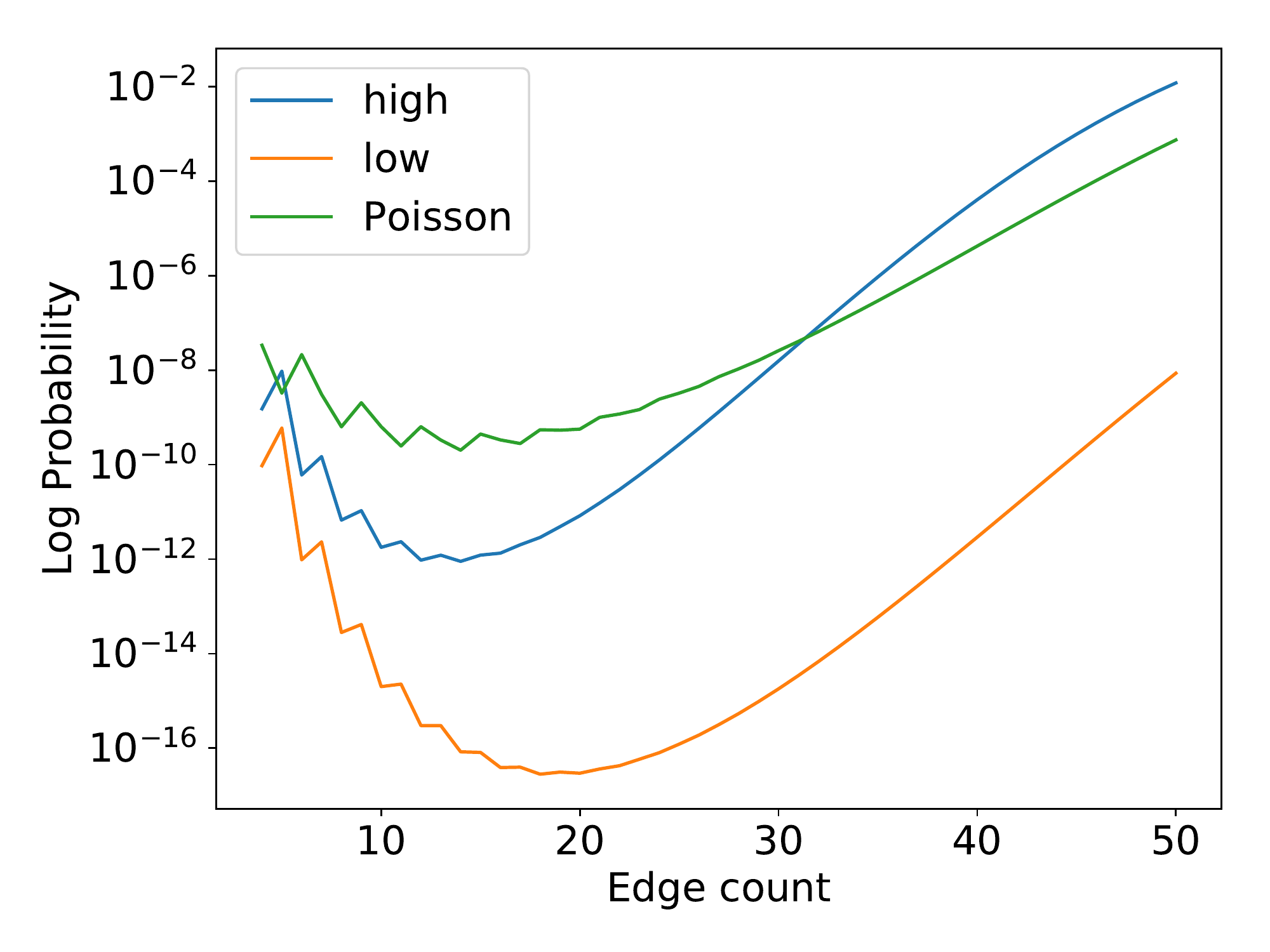}
\end{subfigure}
\caption{Predicted probabilities of local $r$-core formation where $u=e_u \leq 50$ and $k=3$. The left plot shows the log probability of local 1-core formation using graph connectivity (green), the covering heuristic (orange), and the mean of 1M Monte Carlo trials per point (blue). The right plot shows the log probability of local 2-core formation using the upper and lower connectivity bounds (blue and orange) as well as the covering heuristic (green).}
\label{fig:local_cores}
\end{figure}

\subsection{Local $r$-core probabilities}

We first evaluate various techniques for computing $\mathcal{C}_{\bar{u}}(v,p,k,r)$, the probability that an $r$-core forms on a specific set of $u$ vertices in $\mathcal{H}^k_{v, p}$.

Figure~\ref{fig:local_cores} (left) shows the probability of local 1-core formation in 3-uniform hypergraphs with core size $u$ varying from 1 to 50 and having $e_u = u$ expected edges in any set of $u$ vertices. The blue curve shows the result of 1M MC trials per point, which we generated by creating random hypergraphs and testing for the presence of a 1-core. The green curve shows the same probability as determined by Theorem~\ref{thm:conn_prob}, which we call the \emph{connectivity} approach. Finally, the orange curve gives an estimate of 1-core probability using the covering heuristic defined by Equation~\ref{eqn:covering}.

For all values $e_u \in [1, 50]$, the connectivity calculation of 1-core probability closely match that of the MC trials. In contrast, the covering heuristic provides a much less accurate value for 1-core probability when $e_u$ is small. However, as $e_u$ approaches 50, the heuristic becomes much tighter. Not shown in the plots is the numerical breakdown of the connectivity approach. For example, when $e_u=200$, MC trials indicate that the probability that a 1-core forms is 1.9e-4. The connectivity approach predicts -1.84e+23, and the covering heuristic predicts a probability of 2.2e-4. Thus for large values of $e_u$, the connectivity approach breaks down entirely, while the covering heuristic maintains a reasonable estimate. As we will in Section~\ref{sec:global_eval}, the numerical instability of the connectivity calculation propagates to global $r$-core bounds provided by Theorem~\ref{thm:upper_lower}. 

Figure~\ref{fig:local_cores} (right) gives probabilities of local 2-core formation in 3-uniform hypergraphs, where again, core size $u$ varies from 1 to 50 and there are $e_u = u$ edges expected in each subset of $u$ vertices. Here we see upper and lower bounds provided by Theorem~\ref{thm:upper_lower} shown in the blue and orange curves, respectively. The green curve shows an estimate of 2-core probability using the covering heuristic defined by Equation~\ref{eqn:covering}. Overall, the upper and lower bounds are initially close, but diverge significantly for $e_u$ close to 20, and then gradually begin to converge again for larger $e_u$. The covering heuristic initially provides an upper bound on 2-core probability, but for $e_u > 30$, it settles somewhat below the upper bound. 

\begin{figure}
\centering
\begin{subfigure}{.5\textwidth}
  \centering
  \includegraphics[width=0.98\linewidth]{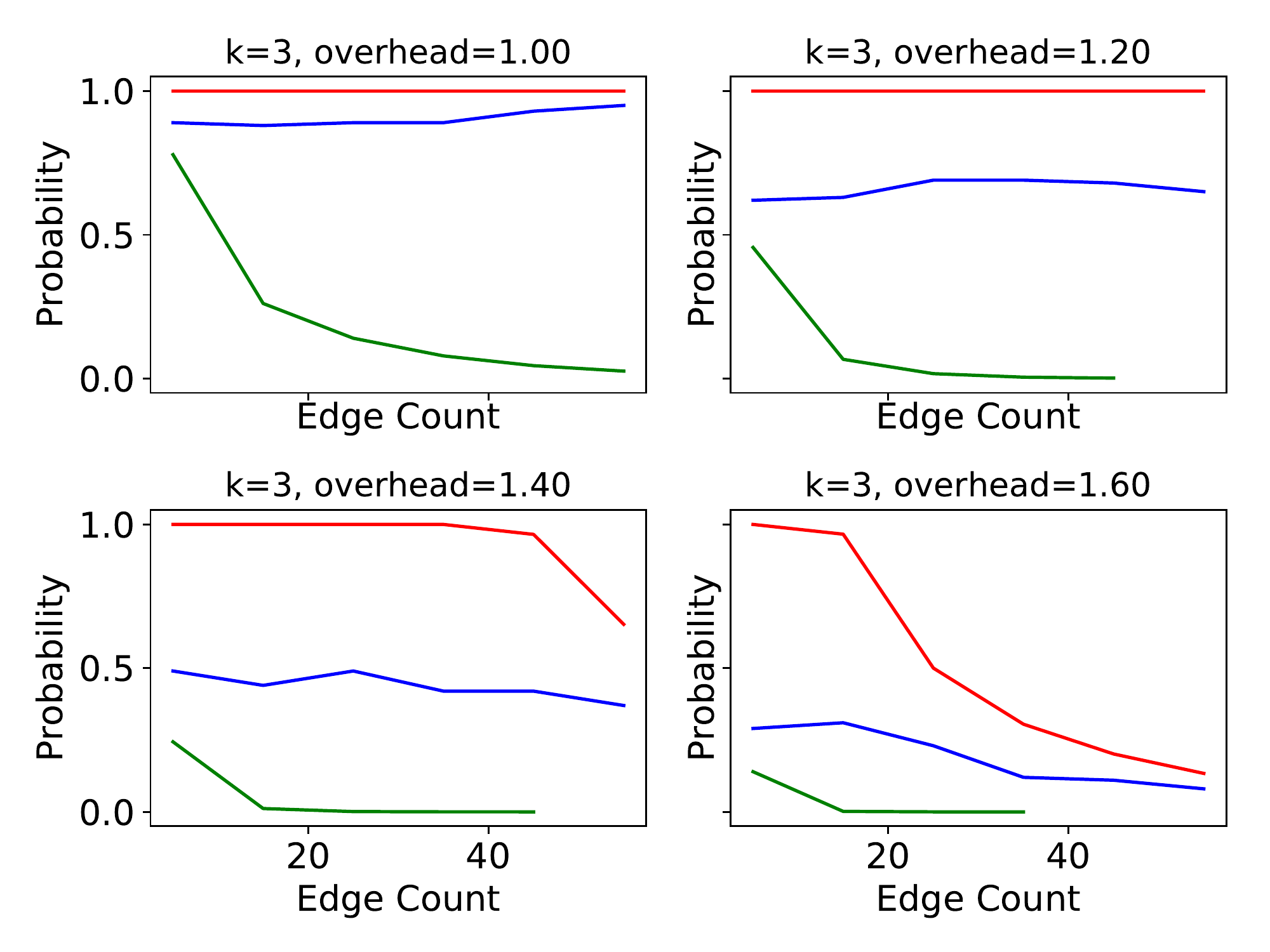}
\end{subfigure}%
\begin{subfigure}{.5\textwidth}
  \centering
  \includegraphics[width=0.98\linewidth]{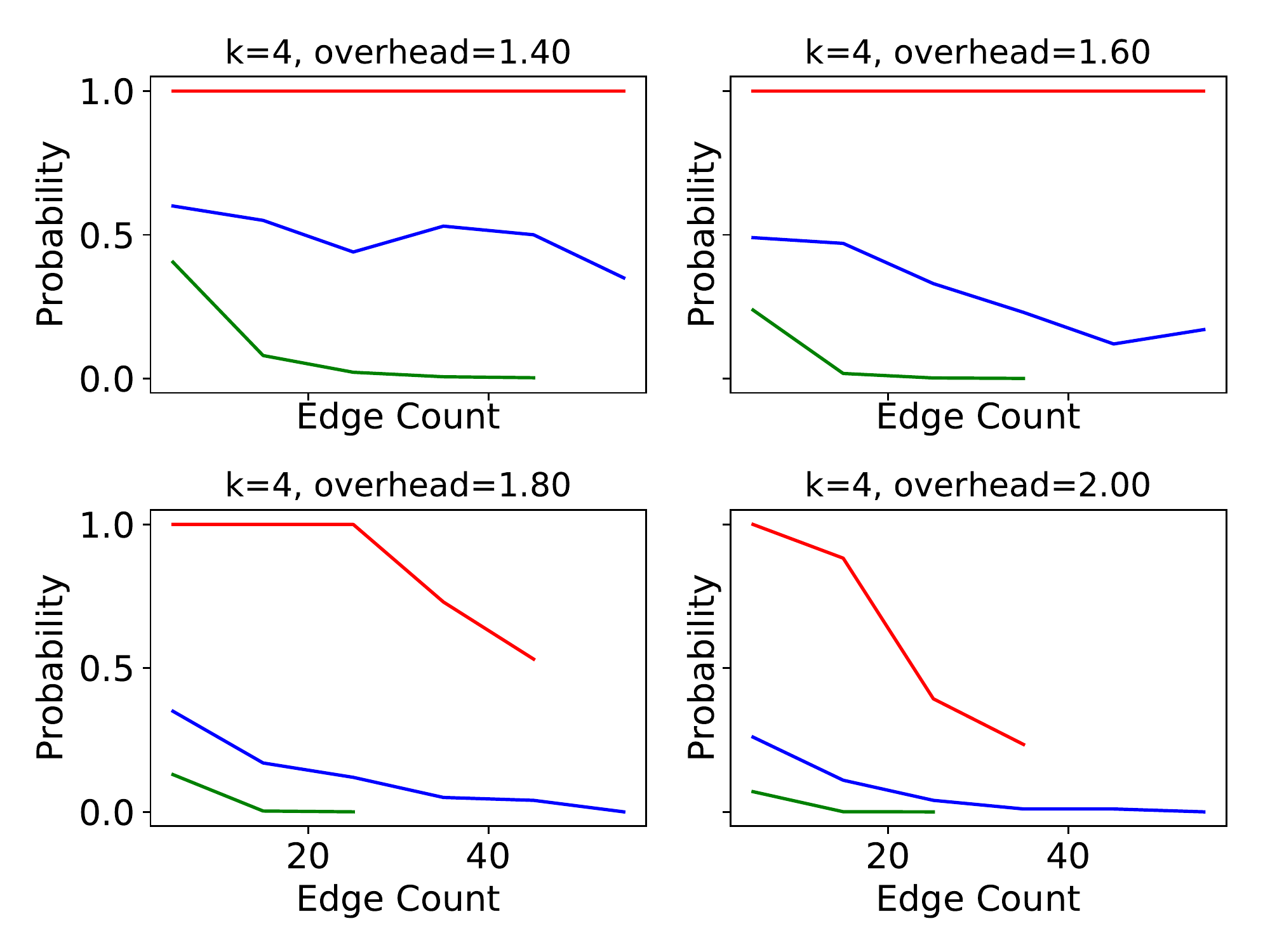}
\end{subfigure}
\caption{Bounds on global $r$-core formation. The left set of plots show the probability of global 2-core formation over 3-uniform hypergraphs (i.e. $k=3$) using graph connectivity (green), the covering heuristic (orange), and the mean of 100 Monte Carlo trials per point (blue). The right set of plots shows the same curves for probability of 2-core formation over 4-uniform hypergraphs (i.e. $k=4$).}
\label{fig:global_cores}
\end{figure}

\subsection{Bounding global $r$-core probabilities}
\label{sec:global_eval}

We next evaluate our bound on $\mathcal{C}(v,p,k,r)$, the probability that at least one $r$-core forms somewhere in $\mathcal{H}^k_{v, p}$.

Figure~\ref{fig:global_cores} shows the probability of 2-core formation anywhere in a 3-uniform (left) or 4-uniform (right) hypergraph. Here we vary the total number of vertices in the hypergraph $v$ and the expected number of edges $e_v$ (both along the independent axis) as well as \emph{overhead}, which is the ratio of vertices to edges (a distinct value for each facet). In particular, if the overhead is $x$ in a given facet, then $e_v$ varies from $k$ up to 60 along the independent axis and $v = x e_v$. The blue curve shows the mean of 100 MC trials per point, while the red and green curves show upper and lower bounds, respectively, using Theorem~\ref{thm:conn_prob} along with Theorem~\ref{thm:upper_lower}. Breaks in the bounds occur after the value $e_v$ where numerical failure is detected; because probability is typically (though perhaps not necessarily) non-increasing with vertex count, and $v$ is a function of $e_v$, we assume numerical breakdown when the probability begins to increase with $e_v$.

There are several notable trends in the plots. First, the bounds become tighter as both $e_v$ and the overhead increase (i.e. as $v$ grows large relative to $e_v$). When $k=3$ and overhead is 1.6, we see that the upper bound drops very close to the MC curve as $e_v$ increases. Similarly, for $k=4$ and overhead equal to 2.0, the MC curve falls closely in-line with the lower bound as $e_v$ grows. Second, numerical instability also appears to increase with overhead. Although some instability is apparent in nearly all plots, it occurs for lower and lower values of $e_v$ as the overhead increases.

\begin{figure}
\centering
\begin{subfigure}{.5\textwidth}
  \centering
  \includegraphics[width=0.98\linewidth]{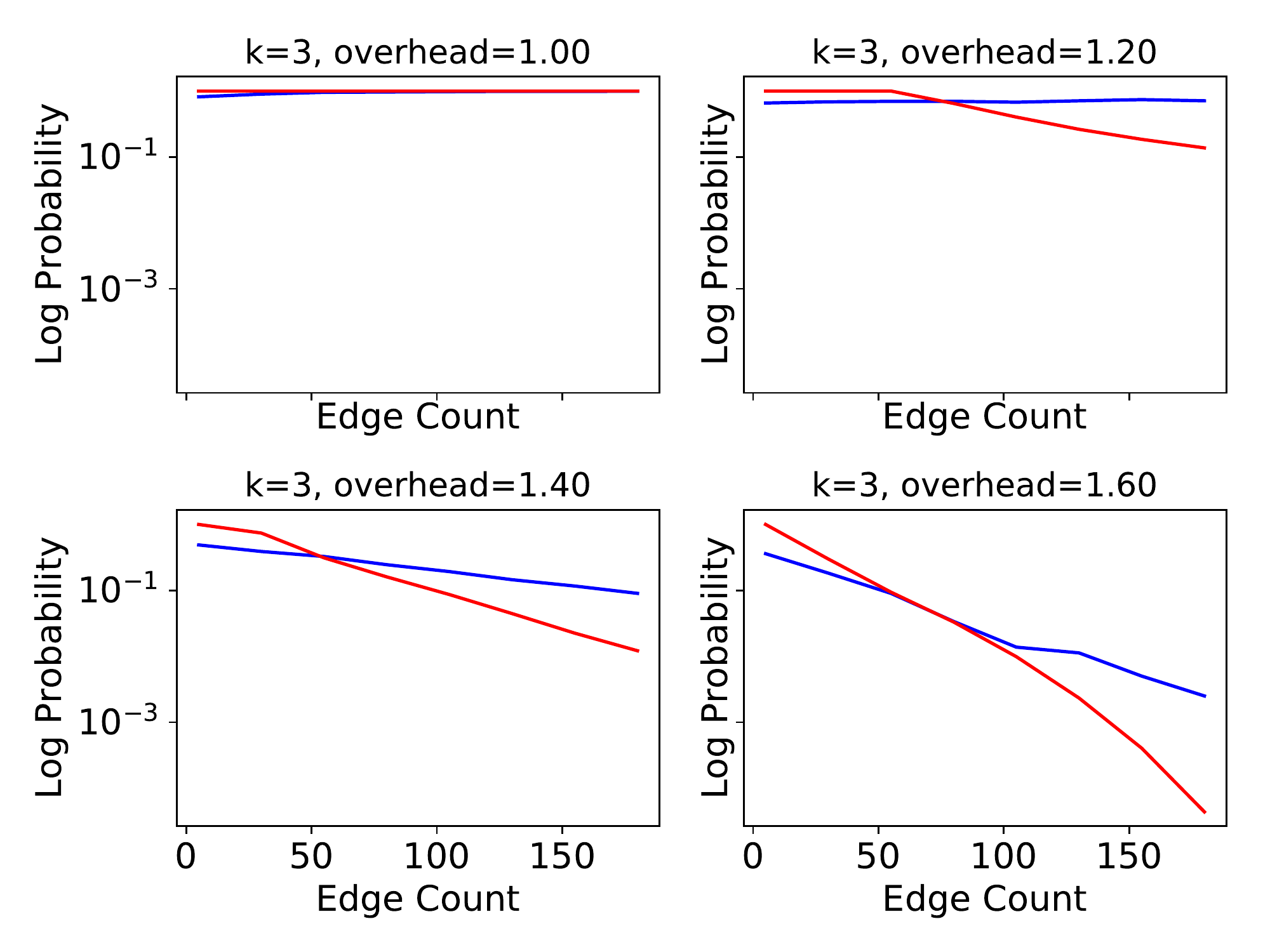}
\end{subfigure}%
\begin{subfigure}{.5\textwidth}
  \centering
  \includegraphics[width=0.98\linewidth]{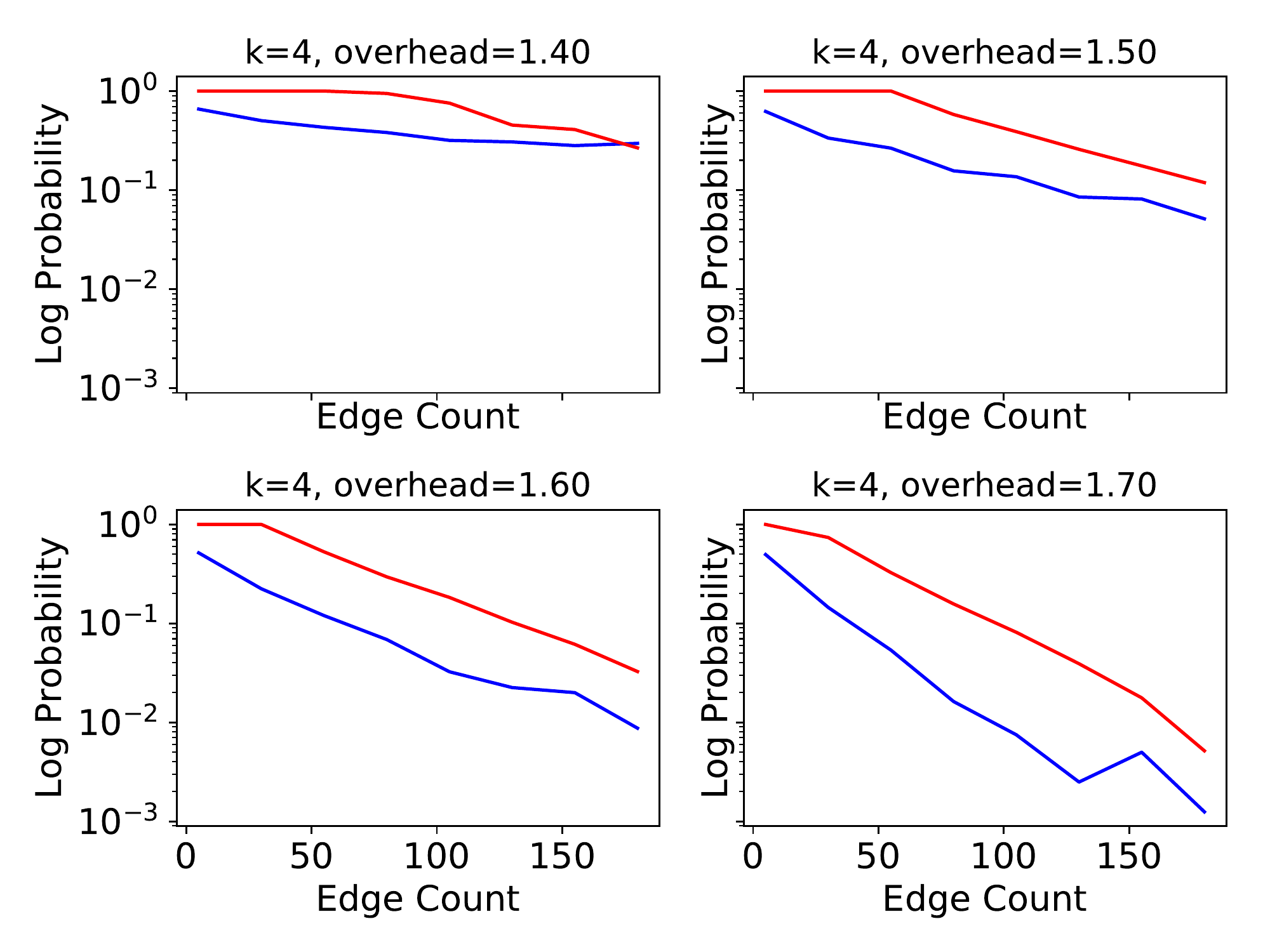}
\end{subfigure}
\caption{Heuristic assessment of global $r$-core formation. The left set of plots show the log probability of global 2-core formation over 3-uniform hypergraphs (i.e. $k=3$) using the Poisson covering heuristic (red), and the mean of 800 Monte Carlo trials per point (blue). The right set of plots shows the same curves for log probability of 2-core formation over 4-uniform hypergraphs (i.e. $k=4$).}
\label{fig:global_cores_bin}
\end{figure}

\subsubsection{Approximate solution}

Due to the numerical breakdown of the connectivity approach, we also explore an approximation to $\mathcal{C}(v,p,k,r)$ that uses the covering heuristic along with Theorem~\ref{thm:upper_lower}. Figure~\ref{fig:global_cores_bin} shows the probability of 2-core formation anywhere in a 3-uniform (left) or 4-uniform (right) hypergraph. Again, we vary the total number of vertices in the hypergraph $v$ and the expected number of edges $e_v$ (both along the independent axis) as well as overhead (a distinct value for each facet). The blue curve shows the mean of 800 MC trials per point. The red curve shows the approximation. 

From the plots we see two major trends. First, the approximation appears to remain above the actual probability when $e_v$ is small, at some point crosses below the probability, and then remains below as $e_v$ continues to increase. Second, the tendency for the approximation to remain above the actual probability appears to increase with $k$. Not shown in the plot are the values for MC and approximate probability when $k=3$, $e_v=500$, overhead was $1.6$, which were 1.67e-3 and 6.28e-9, respectively. Therefore, it appears that the relative difference between actual and approximate probabilities widens considerably as $e_v$ increases.

\section{Conclusion}

We have presented an algorithm for computing the exact probability that a single $r$-core forms in a $k$-regular hypergraph. It can be easily extended to provide an upper bound on the probability that at least one $r$-core forms in the hypergraph. The algorithm requires a subroutine that calculates the probability that a 1-core forms in the induced hypergraph on any given subset of vertices. To that end, we also presented two methods for calculating local 1-core probability. The first method is an exact solution that uses hypergraph connectivity. We prove that this connectivity approach can be used to produce upper and lower bounds on the probability of global $r$-core formation. The second method is an approximation that uses a covering heuristic. The exact solution is shown experimentally to break down numerically for modestly large numbers of vertices (30-50, depending on other parameters). The approximation remains numerically stable and is reasonably accurate for hypergraphs with fewer than 100 vertices; but it is not reliable as either an upper or lower bound.

\urlstyle{sf}
\pagestyle{plain}

{\footnotesize \bibliographystyle{acm}
\bibliography{refs}}

\begin{thebibliography}{1}

\bibitem{Gilbert:1959}
{\sc Gilbert, E.~N.}
\newblock {Random Graphs}.
\newblock In {\em The Annals of Mathematical Statistics\/} (1959), vol.~30,
  pp.~1141--1144.

\bibitem{goodrich:2011}
{\sc Goodrich, M., and Mitzenmacher, M.}
\newblock Invertible bloom lookup tables.
\newblock In {\em 49th Annual Allerton Conference on Communication, Control,
  and Computing\/} (Sept 2011), pp.~792--799.

\bibitem{Molloy:2004}
{\sc Molloy, M.}
\newblock The pure literal rule threshold and cores in random hypergraphs.
\newblock In {\em Proceedings of the Fifteenth Annual ACM-SIAM Symposium on
  Discrete Algorithms\/} (Philadelphia, PA, USA, 2004), SODA '04, Society for
  Industrial and Applied Mathematics, pp.~672--681.

\end{thebibliography}

\appendix

\pagebreak
\section{Notation}
\label{sec:notation}

\begin{center}
\begin{tabular}{ r l }
$r$ & order of hypergraph core \\
$v$ & number of hypergraph vertices  \\ 
$e$ & expected number of hypergraph edges  \\  
$k$  & number of vertices per edge \\
$p$ & the probability that any given edge forms in the hypergraph \\
$b(x; n, p)$ & probability mass function of the distribution $\texttt{Binomial}[ n,p ]$ \\
$B(x; n, p)$ & CDF of the distribution $\texttt{Binomial}[n, p]$ \\
$\mathcal{H}^k_{v, p}$ & a $k$-uniform hypergraph with $v$ vertices and edge probability $p$ \\
$\mathcal{H}^k_{v, p}(U)$ & the induced hypergraph on vertices $U$ \\
$\mathcal{I}^k_{v, p}$ & a $k$-uniform \emph{interleaved} hypergraph with $v$ vertices and edge probability $p$ \\
$\mathcal{C}(v,p,k,r)$ & probability that one or more $r$-cores form \emph{anywhere} in $\mathcal{H}^k_{v, p}$ \\
$\mathcal{C}_*(v,p,k,r)$ & probability that exactly one $r$-core forms \emph{anywhere} in $\mathcal{H}^k_{v, p}$ \\
$\mathcal{C}_u(v,p,k,r)$ & probability that an $r$-core of size $u$ forms \emph{anywhere} in $\mathcal{H}^k_{v, p}$ \\
$\mathcal{C}_{\bar{u}}(v,p,k,r)$ & probability that an $r$-core forms \emph{on a specific set of} $u$ vertices in $\mathcal{H}^k_{v, p}$ \\
$\mathcal{C}_U(v,p,k,r)$ & probability that an $r$-core forms in $\mathcal{H}^k_{v, p}$ on vertices $ U \subseteq V$ \\
$\mathcal{C}^*_U(v,p,k,r)$ & probability that an $r$-core forms on vertices $ U \subseteq V$ in hypergraph $\mathcal{I}^k_{v, p}$ \\
$\tilde{\mathcal{C}}_U(v,p,k,r)$ & Poisson formula approximating $\mathcal{C}_U(v,p,k,r)$ \\
$f^k_p (u)$ & probability that certain $u$ vertices are connected in $\mathcal{H}^k_{v, p}$ \\
$f^k_p (u,i)$ & probability there exists at least one component on $i$ vertices in $\mathcal{H}^k_{u, p}$ \\
$\varepsilon^k(u,i)$ & number of possible edges connecting vertex sets of size $i$ and $u-i$ in $\mathcal{H}^k_{v, p}$
\end{tabular}
\end{center}

\pagebreak
\section{General Algorithm}
\label{sec:gen_alg}

\begin{algorithm}
\caption{Probability exactly one $r$-core forms somewhere in $\mathcal{H}^k_{v, p}$}\label{alg:general}
\begin{algorithmic}[1]
\Procedure{$\mathcal{C}_*$}{$v,p,k,r$}

\State ProbMap = \{\} 

\For{$v' \in \{k, \ldots, v\} $} 
	\State $n' = p \binom{v'}{k}$
	\State ProbMap$[v']$ = \{\}
	\State ProbMap$[v'][v'] = \mathcal{C}_{\bar{v}'}(v',p,k,r)$
	\For{$c \in \{v', \ldots, k\} $} \Comment{loop in reverse order}
		\State ProbGlobalMap$[v'][c] = \binom{m'}{c}$ \Comment{number of subsets of size $c$} 
		\State ProbGlobalMap$[v'][c] \mathrel{*}= \mathcal{C}_{\bar{c}}(c,p,k,r)$ 

		\For{$x \in \{c+1, v'\}$} \Comment{$r$-core not subset of any others}
			\State ProbNoCore = $\texttt{pow}\left(1 - \text{ProbGlobalMap}[v'][x], \binom{v' - c}{x - c}\right)$
			\State $\text{ProbGlobalMap}[v'][c]$ $\mathrel{*}= \text{ProbNoCore}$
		\EndFor
            
		\If{$v' - c > k$} \Comment{exist no other $r$-cores} 
			\State Prob = 0
			\For{$i \in \{\}$} \Comment{add up entries}
				\State Prob $\mathrel{+}= \text{ProbGlobalMap}[x][i]$
			\EndFor
			\State $\text{ProbGlobalMap}[v'][c] \mathrel{*}= 1 - \text{Prob}$ 
		\EndIf
		
	\EndFor
 \EndFor

\State FinalProb = 0
\For{$i \in \{\}$}
	\State FinalProb $\mathrel{+}= \text{ProbGlobalMap}[v][i]$
\EndFor

\State \textbf{return} FinalProb 

\EndProcedure
\end{algorithmic}
\end{algorithm}

\end{document}